# EASM: Efficiency-Aware Switch Migration for Balancing Controller Loads in Software-Defined Networking

[1]Tao Hu, [1]Julong Lan, [1]Jianhui Zhang, [1]Wei Zhao

[1]National Digital Switching System Engineering & Technology Research Center, Zhengzhou, 450002, China

E-mail: hutaondsc@163.com; Telephone number: +86-182-3715-2418

**Abstract**: Distributed multi-controller deployment is a promising method to achieve a scalable and reliable control plane of Software-Defined Networking (SDN). However, it brings a new challenge for balancing loads on the distributed controllers as the network traffic dynamically changes. The unbalanced load distribution on the controllers will increase response delay for processing flows and reduce the controllers' throughput. Switch migration is an effective approach to solve the problem. However, existing schemes focus only on the load balancing performance but ignore migration efficiency, which may result in high migration costs and unnecessary control overheads. This paper proposes Efficiency-Aware Switch Migration (EASM) to balance the controllers' loads and improve migration efficiency. We introduce load difference matrix and trigger factor to measure load balancing on controllers. We also introduce the migration efficiency problem, which considers load balancing rate and migration cost simultaneously to optimally migrate switches. We propose EASM to efficiently solve to the problem. The simulation results show that EASM outperforms baseline schemes by reducing the controller response time by about 21.9%, improving the controller throughput by 30.4% on average, maintaining good load balancing rate, low migration costs and migration time, when the network scale changes.

## 1. Introduction

Software-Defined Networking (SDN) is a new network architecture that decouples data plane from control plan and manages network with a global view [1]. Due to its centralized control, open interface and network programmability, SDN can improve network performance and has been deployed in the Internet and data center networks [2]. However, a single SDN controller has not enough control ability to process increasing number of traffic flows and applications in large-scale networks [3]. Researches in [4]-[6] propose to achieve the logically centralized controller with physically distributed multiple controllers to improve the scalability and reliability of the control plane. Specifically, a network is partitioned into several domains, each of which has one domain controller to manage switches and flow requests [7]. Controllers communicate with each other about domain information to ensure a consistent network view. As traffic varies in the network [8], controllers in different domains could handle the different number of flow requests, and the static matches between switches and controllers may result in unbalanced load allocation on controllers: hot controllers with insufficient control capability and cold controllers with low resource utilization [9].

Dynamic switch migration is an elastic control approach to solve the problem of unbalanced load distribution on controllers. It migrates the control of switches from overloaded controllers to underloaded controllers. However, existing schemes only focus on the balancing performance of control load on controllers but ignore migration efficiency, which may lead to high migration costs, increase control overheads, and squander network resources. In this paper, we propose Efficiency-Aware Switch Migration (EASM) to achieve good load balancing performance on the controllers and low migration costs. The main contributions of this paper are summarized as follows:

- We identify the inefficiency migration problem of existing switch migration schemes and use analysis and examples to explain the undesirable results caused by the existing schemes.
- We propose EASM for effective switch migration. EASM consists of three algorithms. EASM-1 calculates trigger factors to measure the load balancing performance of controllers. If the trigger factor exceeds a threshold, EASM-2 selects migrating switches by solving the migration efficiency problem, which characterizes load balancing rate and migration cost simultaneously. EASM-3 changes the mapping relationship between switches and controllers.
- We evaluate the performance of EASM against baseline schemes. The results show that if the controller load imbalance happens, EASM reduces the controller response time for efficient migration by about 21.9%, improves the controller throughput by 30.4% on average, decreases the migration cost and migration time, and gains the better load balancing performance.

The rest of paper is organized as follows. Section 2 illustrates the motivation. Section 3 introduces the overview of EASM strategy. Sections 4 and 5 detail two components of EASM: load balancing judgement and switch migration design. The simulation results are presented and analyzed in Section 6. Section 7 reviews the related work. Section 8 concludes this paper.

## 2. Motivation

Switch migration is usually used for adjusting the distribution of controller loads through migrating the switch from the overloaded controller to the underloaded controller. However, existing migration designs are difficult to realize good load balancing performance and low migration cost. In this section, we illustrate the problem through an example in Fig. 1 and compare existing solution with our scheme.

Fig. 1 shows an SDN using three distributed controllers. In the figure, the network consists of Domain1, Domain2 and Domain3, and each domain has several switches and is controlled by its domain controller. Table I shows the flow arrival rate of each switch at time T. At time T, we use the total flow rates of a domain to represent the controller load of the domain, and use the normalized load variance to represent load balancing rate (*LBR*), as shown in Eq. (1), where $L_i$ is the load of the $i^{th}$ controller, and $\overline{L}$ is the average load of *n* controllers. *LBR* represents the degree of closeness to the ideal load distribution. The higher *LBR*, the more balanced load distribution.

$$LBR = \frac{1}{n} \cdot \frac{\sum_{i=1}^{n}(L_i - \overline{L})}{\sqrt{\sum_{i=1}^{n}(L_i - \overline{L})^2}} \qquad (1)$$

Table I. The flow arrival rate of switches in the network

| Switch | $s_1$ | $s_2$ | $s_3$ | $s_4$ | $s_5$ | $s_6$ | $s_7$ | $s_8$ | $s_9$ |
|---|---|---|---|---|---|---|---|---|---|
| Flow rate (KB/s) | 30 | 30 | 30 | 30 | 30 | 40 | 50 | 30 | 40 |

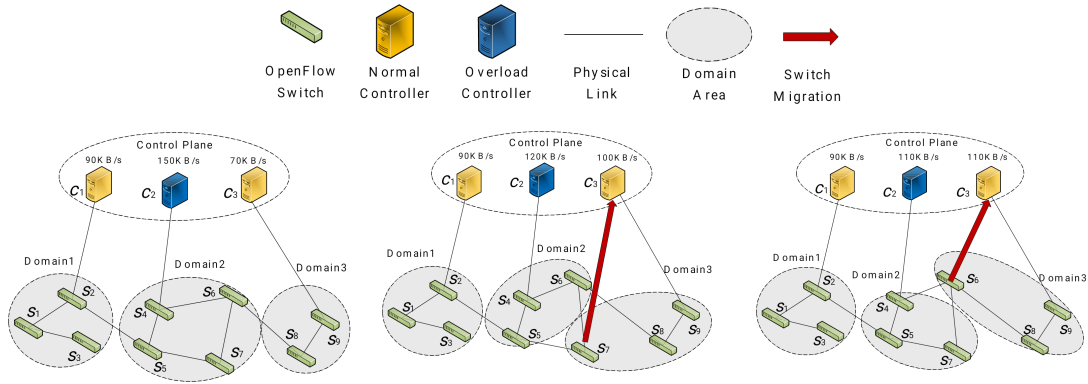

(a) Initial network state under imbalanced control load allocation on controllers (b) Controller load allocation using Existing Switch Migration (c) Controller loads allocation using EASM
Fig. 1. A motivating example for comparison

Table II The comparison of three scenes

|  | Initial network state | Existing Switch Migration | EASM |
|---|---|---|---|
| $c_1$ (KB/s) | 90 | 90 | 90 |
| $c_2$ (KB/s) | 150 | 100 | 110 |
| $c_3$ (KB/s) | 70 | 120 | 110 |
| LBR | 0.641 | 0.852 | 0.915 |
| MC (KB/s) | Null | 200 | 120 |

In Fig. 1(a), the controllers are initialed with the unbalanced loads, and the load balancing rates of three domains are computed as follows.

### Initial network state under Load Imbalance (LI):

$Load_{LI}(c_1) = 30 + 30 + 30 = 90\ KB/s$

$Load_{LI}(c_2) = 30 + 30 + 40 + 50 = 150\ KB/s$

$Load_{LI}(c_3) = 30 + 40 = 70\ KB/s$

$LBR_{LI} = 0.641$

In Fig. 1(a), $c_1$ and $c_3$ are low-utilized controllers, while $c_2$ is an overloaded controller. Existing Switch Migration follows OpenFlow 1.3 [10], where one controller has three roles: master, equal and slave. Master controller is used for processing Packet-in requests sent from switches; equal and slave controllers are used as backup. Each switch connects to one master controller and several slave controllers.

Fig. 1(b) shows the result of using an Existing Switch Migration scheme [11]. In the figure, the load of $c_3$ is the lightest and the flow rate of $s_7$ reaches 50KB/s, ESM migrates $s_7$, the switch with the highest flow from $c_2$ to $c_3$ the lightest loaded controller, to balance controller loads. After migration completed, the controller loads and load balancing rate are updated as follows.

### Existing Switch Migration (ESM)

$Load_{ESM}(c_1) = 30 + 30 + 30 = 90\ KB/s$

$Load_{ESM}(c_2) = 30 + 30 + 40 = 100\ KB/s$

$Load_{ESM}(c_3) = 30 + 40 + 50 = 120\ KB/s$

$LBR_{ESM} = 0.852$

During the switch migration, the network will produce the relevant migration costs during switch migration. We use the product of the flow rate and hop to approximately express migration cost *MC*.

$MC_{ESM} = 50 \times 4 = 200\ KB/s$

Existing Switch Migration brings about the high migration costs, which will aggravate the burden of the overloaded controller. The bigger value of MC, the lower controller throughput. If we can consider switch migration from the perspectives of both load balancing rate and migration costs, the controller performance will be better.

Fig 1.c shows the migration result of EASM. In the figure, by simultaneously considering the load balancing rate and the migration cost, EASM migrates $s_6$, the switch with both higher flow rate and shorter migration hops from $c_2$ to $c_3$, to balance controller loads. After migration completed, the controller loads and load balancing rate are updated as follows.

### Efficiency-aware Switch Migration (EASM)

$Load_{EASM}(c_1) = 30 + 30 + 30 = 90\ KB/s$

$Load_{EASM}(c_2) = 30 + 30 + 50 = 110\ KB/s$

$Load_{EASM}(c_3) = 30 + 40 + 40 = 110\ KB/s$

$LBR_{EASM} = 0.915$

$MC_{EASM} = 40 \times 3 = 120\ KB/s$

Table II show the comparison of three scenes. Compared with ESM, EASM scheme not only improves the load balancing rate, but also reduces the migration costs. From the example, we can see that both load balancing rate and migration cost must be jointly considered in switch migration in order to improve the performance of the controller. We solve three problems in this paper: (i) determining which switches should be migrated; (ii) building the migration efficiency model to select migration switches and controllers; (iii) efficiently implementing switch migration.

## 3. EASM Overview

In this section, we first introduce the notations in this paper and then describe the design of EASM.

### 3.1 Notations

We formulate the SDN network as the undirected graph $G = (V, E)$, where $V$ and $E$ are node set and link set, respectively. We assume all controllers could be deployed in the topology optimally [13], and each controller manages some switches. The primary notations used in this paper are listed in Table III.

Table III Notations

| Notation | Definition |
|---|---|
| $V$ | The set of nodes |
| $E$ | The set of links |
| $M$ | The number of controllers |
| $N$ | The number of switches |
| $|V|$ | The number of nodes |
| $D = \{D_1, ..., D_M\}$ | The domain set |
| $C = \{c_1, ..., c_M\}$ | The controller set |
| $S = \{s_1, ..., s_N\}$ | The switch set |
| $\Omega_m$ | The processing capacity of $c_m$ |
| $\Gamma(c_m)$ | The switch set supervised by $c_m$ |
| $h_{ij}$ | The hop between $s_i$ and $c_j$ |
| $J(s_i, c_m)$ | $J(s_i, c_m) = 1$, if $s_i$ connects to $c_m$ $J(s_i, c_m) = 0$, otherwise |

### 3.2 EASM strategy

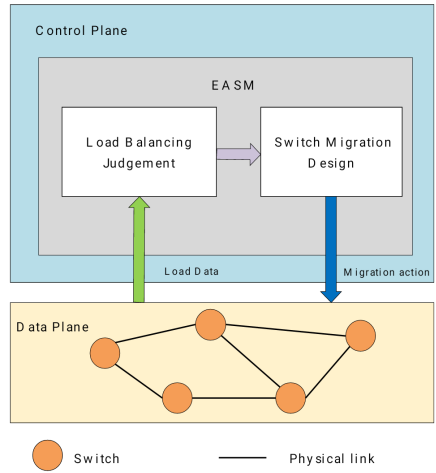

Fig. 2. The overall design of EASM

EASM implements switch migration from the perspective of migration efficiency to improve load balancing rate and reduce migration costs. Fig. 2 shows the overall design of EASM. In the figure, EASM consists of two modules: (1) load balancing judgement; (2) switch migration design. In the load balancing judgment module, EASM measures the controller loads and builds the load difference matrix, and the trigger factor is defined to judge whether the controller loads are balanced. In the switch migration design module, according to migration efficiency model, EASM builds the migration mapping with three factors: emigration controller, migrating switch, and immigration controller, and implements efficient switch migration. Next, we will present the details of load balancing judgement and switch migration design in section 4 and section 5, respectively.

## 4. Load Balancing Judgement

In this section, we will firstly compute the controller's load through synthesizing different network overheads. Then, we introduce a new effective mechanism to judge controller load balancing and design a load imbalance algorithm.

### 4.1 Controller's load

In SDN, the controller loads mainly come from three parts: data interaction overhead for traffic transmission, routing path installment for new flows and state synchronization overhead for global view among controllers.

**Data interaction overhead.** To achieve the centralized control, domain controllers send/receive information from/to switches, including flow table and traffic data of each switch. We formulate data interaction overhead of controller $c_m$ is $F_{data}(c_m)$ as follows:

$$F_{data}(c_m) = \nu \cdot \sum_{s_i \in E(c_m)} h_{im} \cdot J(s_i, c_m) \quad (2)$$

where $\nu$ is the average rate of polling one switch, which depends on the number of links; $h_{im}$ is the hop between $s_i$ and $c_m$; $J(s_i, c_m)$ is the connection relationship of $s_i$ and $c_m$.

**Routing formulation overhead.** When a switch receives a new flow, it sends Packet-in requests to controller and asks for the flow's routing path. Upon receiving the request, the controller calculates a new path for the flow and establishes the path by installing flow entries in switches on the path. Fig. 3 shows the routing formulation of controller. In the figure, a new flow destines to $s_5$ arrives at $s_2$. $s_2$ sends a request to its master controller $c_1$, and $c_1$ must process Packet-in sent by switches and establishes the path.

Therefore, for $c_m$, routing formulation overhead $F_{routing}(c_m)$ contains two parts that are Packet-in processing $f_{packet}(c_m)$ and flow table distributing $f_{table}(c_m)$,

$$f_{packet}(c_m) = P_{packet} \cdot \sum_{s_i \in S} \sum_{c_m \in C} h_{im} \cdot J(s_i, c_m) \quad (3)$$

$$f_{table}(c_m) = \sum_{s_i \in S} \sum_{c_n, c_m \in C} \alpha_{s_i} \cdot h_{mn} \cdot h_{im} \cdot J(c_m, c_n) \quad (4)$$

$$F_{routing}(c_m) = f_{packet}(c_m) + f_{table}(c_m) \quad (5)$$

where $P_{packet}$ is the average size of Packet-in sent by switch; $\alpha_{s_i}$ is the average flow rate of switch $s_i$; $h_{im}$ is the hop between $s_i$ and $c_m$; $J(c_m, c_n)$ is the connection relationship of $c_m$ and $c_n$.

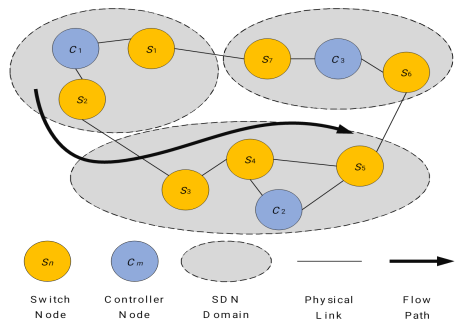

Fig. 3. Routing formulation in the distributed network

**State synchronization overhead.** In the multi-controller SDN network, synchronization messages are sent between controllers to maintain the global network view, producing state synchronization overhead $F_{state}(c_m)$,

$$F_{state}(c_m) = \zeta_{sync} \cdot \sum_{c_m, c_n \in C} J(c_m, c_n) \cdot h_{mn} \tag{6}$$

where $\zeta_{sync}$ is the size of synchronization packet; $J(c_m, c_n)$ is the connection relationship of $c_m$ and $c_n$; $h_{mn}$ is the hop between $c_m$ and $c_m$.

Therefore, the controller loads are the linear aggregation of the three overheads in the network [13]. The computation of load $L(c_m)$ is shown in Eq. (7).

$$L(c_m) = \sigma_1 \cdot F_{data}(c_m) + \sigma_2 \cdot F_{routing}(c_m) + \sigma_2 \cdot F_{state}(c_m) \tag{7}$$

$$\sum_{i=1}^{3} \sigma_i = 1 \tag{8}$$

where $\sigma_1$, $\sigma_2$ and $\sigma_3$ are the corresponding weights for different overheads, respectively.

### 4.2 Effective mechanism

We design a simple but effective mechanism to determine whether the controller loads are balanced in the network.

Firstly, we generate the load difference matrix:

$$D_{M \times M} = \begin{Bmatrix} d(c_1, c_1) & d(c_1, c_2) & \ldots & d(c_1, c_M) \\ d(c_2, c_1) & d(c_2, c_2) & \ldots & d(c_2, c_M) \\ \ldots & \ldots & \ldots & \ldots \\ d(c_M, c_1) & d(c_M, c_2) & \ldots & d(c_M, c_M) \end{Bmatrix} \tag{9}$$

where $d(c_m, c_n) = \dfrac{L(c_m)}{L(c_n)}$, which mean the load difference between controller $c_m$ and $c_n$.

For a given load difference matrix, the balancing judgement is shown in Eq. (10),

$$\exists c_m, c_n \in C, \delta_{mn} = |d(c_m, c_n) - d(c_n, c_m)| > \Lambda \tag{10}$$

where $\delta_{mn}$ is the trigger factor. If $\delta_{mn}$ is larger than the threshold $\Lambda$, there is controller load imbalance in the network, and we need to carry out switch migration at the moment.

Equation (11) shows the computation of the threshold,

$$\Lambda = \frac{maxD_{M \times M} - minD_{M \times M}}{maxD_{M \times M}} \tag{11}$$

where $maxD_{M \times M}$ and $minD_{M \times M}$ represent the maximum load difference and the minimum load difference, respectively.

**Example**. Based on the load imbalance scenario in Fig. 1(a), we use an example to illustrate the validity of our proposed balancing judgement mechanism.

In Fig. 1(a), $L_{LI}(c_1) = 90\ KB/s$, $L_{LI}(c_2) = 150\ KB/s$, and $L_{LI}(c_3) = 70\ KB/s$. Thus, we can get the load difference matrix:

$$D_{3 \times 3} = \begin{bmatrix} d(c_1, c_1) & d(c_1, c_2) & d(c_1, c_3) \\ d(c_2, c_1) & d(c_2, c_2) & d(c_2, c_3) \\ d(c_3, c_1) & d(c_3, c_2) & d(c_3, c_3) \end{bmatrix} = \begin{bmatrix} 1.0 & 0.6 & 1.3 \\ 1.7 & 1.0 & 2.1 \\ 0.9 & 0.5 & 1.0 \end{bmatrix}$$

We get $\delta_{21} = 1.1$, $\delta_{23} = 1.6$, $\delta_{13} = 0.4$ and $\Lambda = 0.7$, respectively. Because both $\delta_{21}$ and $\delta_{23}$ are larger than $\Lambda$, we identify that there is load imbalance. Using the method in Fig. 1(b) and Fig.1 (c) also show the same result. We can see that our balancing judgement mechanism is valid.

#### 4.2.1 Load imbalance detection algorithm

We design load imbalance detection algorithm for this mechanism, which is described as follows. In SDN network, controllers interact information with switches, and compute the aggregated load $L(c_m)$ and load difference $d(c_m, c_n)$ (Line 1). We get the load difference matrix $D_{M \times M}$ at the moment (Line 2). Then we compute the trigger factor $\delta_{mn}$ for different controllers in $D_{M \times M}$, and compare it with threshold $\Lambda$ (Line 5). If $\delta_{mn}$ surpasses this threshold, we conclude that there is load imbalance in the network (Line 6). All trigger factors, which surpass the threshold, will be added into a new set $TF$ (Line 7). The pseudo-code of the algorithm is shown in Table IV.

Table IV Load Imbalance Detection

| EASM-1: Load Imbalance Detection |
|---|
| Input:      SDN network $G=(V,E)$ |
| Output:     Trigger factor set $TF$ |
| 1:    For each controller, get $L(c_m)$ and $d(c_m,c_n)$ |
| 2:    Construct matrix $D_{M \times M}$, and compute $\Lambda$ |
| 3:    **while** ($D_{M \times M} = \phi$) |
| 4:      Compute $\delta_{mn}$ |
| 5:      **if** ($\delta_{mn} > \Lambda$) |
| 6:        Detect load imbalance |
| 7:        Add $\delta_{mn}$ to set $TF$ |
| 8:      **endif** |
| 9:      $D_{M \times M} = D_{M \times M} - \{d(c_m,c_n), d(c_n,c_m)\}$ |
| 10:    **endwhile** |

In EASM-1, Line 1 gets the aggregated loads and load difference, and its time complexity is $O(M)$; Line 2 constructs the load difference matrix, and its time complexity is $O(M^2)$; Line 5 computes trigger factor, and its time complexity is $O(N)$; Line 6 to Line11 generates $TF$, and its time complexity is $O(M(M-1))$. Thus, the overall time complexity of EASM-1 is $O(M^2)$.

## 5. Switch Migration Design

In this section, we will determine the migrating objects for switch migration, including emigration controller, migrating switch and immigration controller. Then, we implement dynamic migration decision according to the presupposed migration triplet.

### 5.1 Migrating objects determination

With the help of load difference matrix and trigger factor, we have detected the load imbalance in the network. Next, we will determine migration objects, including emigration controller, migrating switch and immigration controller.

**(1) Emigration controller**

In order to balance controller loads quickly, we set the overloaded controller as the emigration controller. Therefore, according to the characteristic of the constructed load difference matrix, we compute the trigger factors between all controllers. Meanwhile, we get controller $c_m$ and $c_n$ from $\delta_{mn}$ if $\delta_{mn} > \Lambda$. In particular, we assume $L(c_m) > L(c_n)$, and then set $c_m$ as the emigration controller.

Through traversing the entire load difference matrix, we can get several emigration controllers, and all of them will be stored in emigration controller set $C_{EM}$.

**(2) Migrating switch and immigration controller**

Through analyzing and comparing the cases in motivation (Section II), we can find the selections of migration switches and immigration controllers have the big influences in load balancing rate and migration costs. Therefore, we introduce the migration efficiency model, which characterizes the load balancing rate and the migration cost simultaneously, to optimally select the migrating switches the immigration controllers.

Firstly, we give the definitions and computations of the migration cost and load balancing rate.

**Definition 1.** Migration cost. When switch $s_i$ is migrated from controller $c_m$ to $c_n$, it will generate migration cost $MC_{c_m,c_n}^{s_i}$, including migrating request (the front part of Eq. 12) and load change (the latter part of Eq. 12). Next, we will analyze the migration cost in detail.

Migrating a switch from the one controller to another controller produces network cost. We define the consumption of network resource as the migration cost. The migration cost of a switch consists of two parts: (i) migrating request cost and (ii) load change cost. They are

detailed below:

(i) Migrating request cost. During a switch migration, this switch firstly sends a communication packet, which is similar to Packet-in packet, to the immigration controller to request migration. This cost of the procedure is the migrating request cost. Concretely, when switch $s_i$ is migrated from controller $c_m$ to controller $c_n$, the migrating request cost can be computed as $P_{Packet} \cdot \sum J(s_i, c_n)$, where $P_{Packet}$ is the average size of Packet-in sent by the switch, and $J(s_i, c_n)$ is a binary variable that describes the connection relationship between switch $s_i$ and controller $c_n$. The connection relationship firstly appears in Section 3.1. We calculate the value of $J(s_i, c_m)$ based on the physical connection between $s_i$ and $c_m$. $J(s_i, c_m) = 1$ means $s_i$ connects with $c_m$, otherwise $J(s_i, c_m) = 0$.

(ii) Load change cost. If the immigration controller accepts the migrated switch, the switch's traffic will be handled by controller. This process causes the load change of controllers, and the cost of the procedure is load change cost. Here, we consider the controller's load is only related to switch traffic and the length of a path from a switch to the controller. The load change cost can be computed as $\alpha_{s_i} \cdot |h_{in} - h_{im}|$, where $\alpha_{s_i}$ is the average flow rate of switch $s_i$; $h_{in}$ is the length of the path between $s_i$ and $c_n$; $h_{im}$ is the length of the path between $s_i$ and $c_m$.

Based on above analysis, we formulate the migration cost with Eq. 12.

$$MC_{c_m, c_n}^{s_i} = P_{Packet} \cdot \sum J(s_i, c_n) + \alpha_{s_i} \cdot |h_{in} - h_{im}| \tag{12}$$

**Definition 2.** Load balancing rate. This paper computes the controller load variance as load balancing rate, and $\overline{L}$ is the average load of controllers. Before migrating the switch, we get load balancing rate:

$$\eta = \frac{1}{M} \cdot \sum_{m=1}^{M} (L(c_m) - \overline{L})^2 \tag{13}$$

After the switch has migrated, $\eta^*$, $\overline{L}^*$, $L^*(c_m)$ and $L^*(c_n)$ are updated, and the results are shown in Eq. (14) to Eq. (16).

$$\eta^* = \frac{1}{M} \cdot \sum_{m=1, m \neq n}^{M} [(L^*(c_m) - \overline{L}^*)^2 + (L^*(c_n) - \overline{L}^*)^2] \tag{14}$$

$$L^*(c_m) = L(c_m) - \alpha_{s_i} \cdot h_{im} \tag{15}$$

$$L^*(c_n) = L(c_n) + \alpha_{s_i} \cdot h_{in} \tag{16}$$

**Definition 3.** Migration efficiency. We define the ratio of load balancing rate changing and migration cost as the migration efficiency $\tau_{s_i c_n}$. The higher $\tau_{s_i c_n}$, the better controller performance after migration.

$$\tau_{s_i c_n} = \frac{|\eta^* - \eta|}{MC_{c_m, c_n}^{s_i}} \tag{17}$$

$$\forall s_i \in S, c_j \in C, J(s_i, c_j) = \{0, 1\} \tag{18}$$

$$\forall s_i \in S, \sum_{c_m \in C} J(s_i, c_m) = 1 \tag{19}$$

$$\exists c_m \in C, L(c_m) \leq \Omega_m \tag{20}$$

Equation (18) restricts the connections of all devices. Equation (19) represents that each switch only connects with one controller. Equation (20) shows there is no possible that all controllers are in the overloaded states.

Based on the migration efficiency model, we design to select migrating switches and immigration controllers.

**Migrating switch selecting**

The migrating switch $s_i$ is selected from the switch set $\Gamma(c_m)$ managed with emigration controller $c_m$, and $s_i$ must consider the following conditions. First, $c_m$ is more willing to migrate the switch with the high migration efficiency to relief its loads. Moreover, from the perspective of delay, $c_m$ preferentially abandons the switch that is far from it. Therefore, we select the migrating switch based on a probability distribution, which is shown in Eq. (21) and Eq. (22).

$$s_i = \arg\max_{\Gamma(c_m)} \rho_{s_i} \tag{21}$$

$$\rho_{s_i} = \tau_{s_i c_n} \cdot \frac{|\overline{L} - (L(c_m) \cdot |\Gamma(c_m)| - \alpha_{s_i})| \cdot e^{(maxh_{im})}}{\sum_{s_j \in \Gamma(c_m)} e^{(maxh_{im})}} \tag{22}$$

### Immigration controller selecting

When the migrating switch $s_i$ is moved into its slave controller, it firstly detects whether the migration will cause a new overloaded controller. If so, this controller will be abandoned. Therefore, under the guidance of the migration efficiency model, $c_n$ will be selected as the immigration controller according to Eq. (23) and Eq. (24).

$$c_n = \arg\max\{\Phi_n\} \tag{23}$$

$$\Phi_n = \gamma \cdot [\Omega_n - L(c_n) - \alpha_{s_i}] + (1-\gamma) \cdot \tau_{s_i c_n} \tag{24}$$

where $\Phi_n$ represents the weighted sum of remaining processing capacity and the migration efficiency, and $\gamma$ is the corresponding weight.

Based on the above computation, we can get several immigration controllers, and they are saved in immigration controller set $C_{IM}$.

### 5.1.1 Optimal object selection algorithm

Based on the known loads condition, we will select the optimal migration objects in EASM-2. Firstly, for any $\delta_{mn}$ in $TF$, if $L(c_m) > L(c_n)$, we set $c_m$ as the emigration controller and add it into the set $C_{EM}$ (Line 3 to Line 5). Then, we compute migration costs and load balancing rate to get the migration efficiency (Line 8). The migrating switch is selected according to the maximum selection probability (Line 10). The selection of the immigration controller is optimized by SA method. Initial temperature decreases to a moderate stage until the system comes to a balance point, where no more changes require (Line 16 to Line 18). In the next stage, it begins with a lower temperature and allows the model to move toward the better solution (Line 19 to Line 22). The selected immigration controller will be added into the set $C_{IM}$ (Line 25). All migration objects are determined in the end. The pseudo-code of the algorithm is shown in Table V.

Table V Optimal Object Selection

| **EASM-2: Optimal Object Selection** |
|---|
| Input: Trigger factor set $TF$ |
| Output: Emigration controller set $C_{EM}$ |
|         Migrating switch $s_i$ |
|         Immigration controller set $C_{IM}$ |
| **Procedure** emigration controller selection |
| 1:   **while** ( $TF \neq \phi$ ) |
| 2:       Select $\delta_{mn}$ from $TF$ |
| 3:       **if** $L(c_m) > L(c_n)$ |
| 4:        Add $c_m$ into $C_{EM}$ |
| 5:       **endif** |
| 6:       $TF = TF - \{\delta_{mn}\}$ |
| 7:   **endwhile** |
| |
| **Procedure** migrating switch selection |
| 8:   Get migration efficiency $\tau$ |
| 9:   Compute $\rho_{s_i}$ of switch managed by $c_m$ |
| 10:  migrating switch $s_i = \arg\max \rho_{s_i}$ |
| |
| **Procedure** immigration controller selection |
| 11:  Initial temperature $T_0$, |

```
12:     Get  $\Phi_n(T_0)$
13:     for  $k = Max\_temp\_change$  do
14:         $T_k = T_0/(k+1)$
15:         Apply mutation to current state  $\Phi_n(T_k)$
16:         if  $\Phi_n(T_k)_{mut} > \Phi_n(T_k)$
17:             then  $\Phi_n(T_k) = \Phi_n(T_k)_{mut}$
18:         endif
19:         if  $e^{\Delta T/T_k} > \omega$
20:             then  $\Phi_n(T_k) = \Phi_n(T_k)_{mut}$
21:         else Discard  $\Phi_n(T_k)_{mut}$
22:         endif
23:         k=k+1
24:     endfor
25:     Add  $c_n = \arg max\{\Phi_n\}$   into  $C_{IM}$
```

In EASM-2, Line 1 to Line 8 selects the migrating emigration controller, and its time complexity is $O(2(M-1))$. Line 10 computes the migration efficiency, and its time complexity is $O(M \cdot N)$. The time complexity of selecting the migrating switch is $O(M)$. After Line 11, SA method is implemented for selecting the immigration controller, and its time complexity is $O(M \cdot k)$, which is related to initial temperature and cooling rate. Therefore, the overall time complexity of EASM-2 is $O(M \cdot N)$.

### 5.2 Migration decision formulation

Through determining the migration objects, we have acquired the required elements of switch migration. However, the relationships of migration objects aren't one-to-one correspondence, and one emigration controller may migrate switches into the multiple immigration controllers. Therefore, in order to ensure the well-organized and efficient switch migration, we introduce the triplet to represent the precise mapping between migration objects.

**Definition 4.** Migration Triplet. For any switch migration, the migration triplet is defined as $[c_m, s_i, c_n]$, where those three elements form the determined migration mapping. $c_m$ is the emigration controller, selected from $C_{EM}$; $s_i$ is migrating switch, selected from $\Gamma(c_m)$; $c_n$ is the immigration controller, selected from $C_{IM}$. When $[c_m, s_i, c_n]$ is constructed, $c_m$ have no choice but to migrate $s_i$ to $c_n$.

In practice, there may be multiple switches needed to be migrated in the network, so all triplets form a set $Tr$, which is required to update in real-time after every switch is migrated. By this way, it can prevent the migration disorder effectively. When all migrating switches are migrated into those target controllers, we will redetect whether the controller loads meet the load balancing after the migrations completed. According to the final results, we decide to quit EASM or return module 2 until meeting the requirement of load balancing ($\forall c_m, c_n \in C, \delta_{mn} < \Lambda$).

As a dynamic balancing method, switch migration will cost the particular control resources. Particularly, when the load balancing condition becomes better, we must reduce the occurrence of migration to avoid unnecessary consumption of the controller resources. Therefore, in order to achieve the better migration effects, we can adjust the threshold $\Lambda$ according to Eq. (11) after network update. Moreover, the bigger $\Lambda$, and the lower migration frequency.

Particularly, if there are multiple migrating objects with the same migration efficiency and multiple immigration controllers, we migrate those switches to their closest controllers. For example, the switch will be migrated into the controller with the minimum path length among all controllers. If the path length is same, then EASM would randomly select switch to migrate.

#### 5.2.1 Dynamic migration decision algorithm

In EASM-3. Firstly, we select the elements from the migration object sets to construct the triplet set $Tr$ (Line 2), which includes a series of migration mappings. For each $[c_m, s_i, c_n]$, we will migrate $s_i$ from $c_m$ to $c_n$ (Line4 to Line 5), and then shrink $Tr$ (Line 8). After $Tr$ is empty, we will update the network state and improve the threshold $\Lambda$ to complete the dynamic switch migration (Line 10). The pseudo-code of the algorithm is shown in Table VI.

Table VI Dynamic Migration Decision

| **EASM-3: Dynamic Migration Decision** |
|---|
| Input:    Emigration controller set $C_{EM}$ |
|               Migrating switch $s_i$ |
|               Immigration controller set $C_{IM}$ |
| Output:  New network state |
| 1:   $c_m \leftarrow C_{EM}$ ; $s_i \leftarrow \Gamma(c_m)$ ; $c_n \leftarrow C_{IM}$ |
| 2:   Construct triplet set $Tr$ |
| 3:   **while** ( $Tr \neq \phi$ ) |
| 4:       $[c_m, s_i, c_n] \leftarrow Tr$ |
| 5:       Migrate $s_i$ from $c_m$ to $c_n$ |
| 6:       $\Gamma(c_n) = \Gamma(c_n) + \{s_i\}$ , $\Gamma(c_m) = \Gamma(c_m) - \{s_i\}$ |
| 7:       Update states of $c_m$ and $c_n$ |
| 8:       $Tr = Tr - \{[c_m, s_i, c_n]\}$ |
| 9:   **endwhile** |
| 10: Update $\Lambda$ and network state |

In EASM-3, it mainly executes switch migration and updates controller states, and its time complexity is associated with triplet. The overall time complexity of EASM-3 is $O(M+N)$.

# 6. Evaluation

## 6.1 Simulation setting

In this section, we evaluate the performance of EASM under the experimental environment shown in Fig. 4, and make the following descriptions.

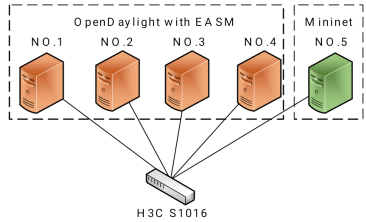

Fig. 4. The experimental topology

(1) Experimental platform

We select OpenDaylight [22] as the experimental controller, and use Mininet [23] as a test platform. OpenDaylight is programmed by Java and supports multiple versions of OpenFlow protocols. Mininet developed by Stanford University is set as the test platform. The physical devices contain five servers with the same configuration (Intel Core i7 3.5GHz 4GB RAM). The operation system is Ubuntu 16.04 and the development kit is JAVA 8. EASM is designed in the application layer of OpenDaylight controller. Considering the performance conflict between OpenDaylight and Mininet on one server, we run OpenDaylight with EASM on four servers (NO. 1- 4) and install Mininet on one server (NO. 5). All servers are connected by H3C S1016 switch.

(2) Topology selecting

We select the authoritative network topology to make the experiments more persuasive. First, we demonstrate the validity of EASM in Internet2 OS3E [24] with 34 nodes and 42 links. Then, we reselect several topologies from Topology Zoo [25] to prove the load balancing performance and the topological adaptability.

(3) Parameters setting

We use Iperf [26] to generate TCP flows to simulate the distribution of the network traffic. The average flow requests are 200KB/s. The controller capacity is limited to 5MB, and $v=15KB/s$, $P_{Packet}=30Byte$, $\zeta_{sync}=18Byte$. The link bandwidth is finite, thus we set the number of switches managed with one controller is from 5 to 20 [27].

(4) Simulation comparison

To verify the performance of EASM, we compare it with the other three strategies.

- **No Switch Migration (NSM)**: the connections between switches and controllers are static.
- **Closest Switch Migration (CSM)**: the overloaded controller randomly migrates the switches into the closest underloaded controller to solve the load imbalance [11].
- **Maximum Utilization Switch Migration (MUSM):** like the typical switch migration scheme, it migrates switch into the controller that has maximum residual capacity [18].
- **Efficiency-aware Switch Migration (EASM):** the switch migration is implemented according to migration efficiency to improve load balancing rate and reduce the migration cost.

The evaluation indexes include controller response time, controller throughput, migration cost and migration time, and load balancing rate.

## 6.2 Result analysis

### 6.2.1 Controller response time

Controller response time is one of evaluation indexes. When the load imbalance occurs, controller response time will be increased significantly. In the experiment, we change the flow request counts to make some controllers overload, and observe the change of controller response time. Flow request count of OS3E is shown in Fig. 5, and each simulation time is 12 hours.

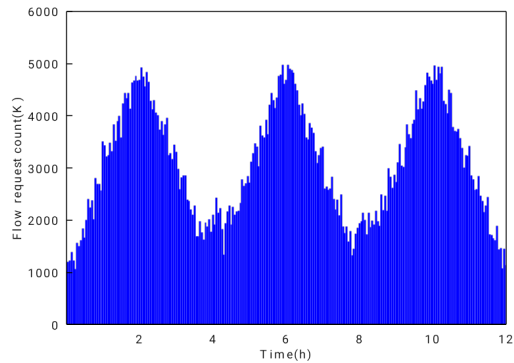

Fig. 5. Flow request counts in OS3E

The average controller response time of four strategies is shown in Fig. 6. We can see that NSM has the most drastic time fluctuation with the change of flow request counts. CSM and MUSM have the smaller fluctuation range, and EASM has the slightest fluctuation. The reasons are explained as follows. Because NSM does not implement switch migration during load imbalance, there is the biggest difference of the controller response time between the normal controller and the overloaded controller. Although CSM and MUSM adopt switch migration to balance controller loads and lower response time, nearest migration is easy to cause new load imbalance after migration, and MUSUM is lack of global planning. EASM analyses the composition of controller loads in detail and constructs the load difference matrix to avoid the local optimal problem, which could ensure the high-efficiency migration and reduce the controller loads quickly. Compared with other strategies, the average controller response time of EASM has reduced 21.9% at least.

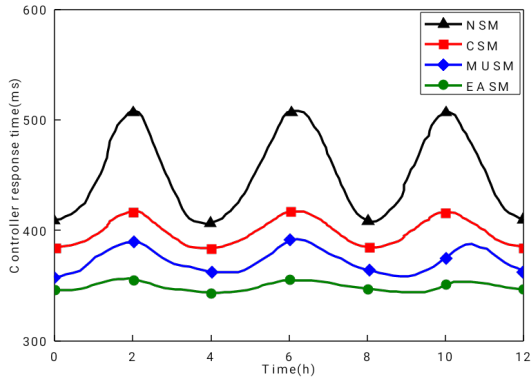
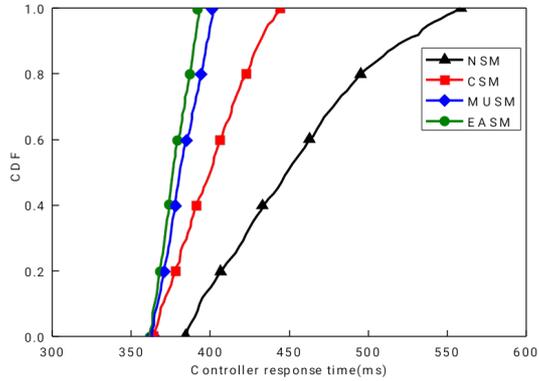

Fig. 6. Average controller response time    Fig. 7. CDFs for different strategies

The cumulative distribution function (CDF) of controller response time is shown in Fig. 7. Due to the reasonable migration model setting, EASM is less vulnerable to the change of flow request counts than other strategies.

### 6.2.2 Controller throughput

Based on the flow request counts in Fig. 5, we use the average controller throughput to reflect the load condition. The higher throughput, the better controller performance. The experiment result is shown in Fig. 8.

Due to the static connection of NSM, it has the lowest throughput, which is less than 2000packets/s. The remaining three strategies implement switch migration during controller overload, so the average controller throughput has been improved obviously. CSM and MUSM have the similar throughputs, which are close to 3000packets/s. Differing from the unilateral migration decision (the shortest distance in CSM and the largest capacity in MUSM), EASM makes efforts to improve load balancing rate while reducing migration cost through setting migration efficiency model. Meanwhile, the reasonable design of triplet also ensures the concurrent and coordinating migrations. Therefore, the average controller throughput of EASM reaches about 3660packets/s, which has increased by 30.4% on average compared with the other migration strategies.

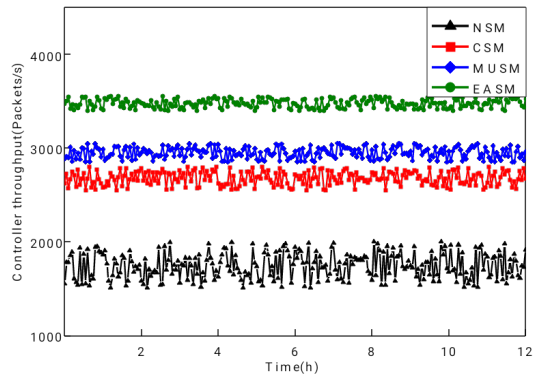

Fig. 8. Controller throughput

### 6.2.3 Migration cost and migration time

In this experiment, we remove NSM and only record the migration costs and time of CSM, MUSM and EASM. That is because NSM does not perform switch migration. As shown in Fig. 9, in terms of migration cost, MUSM is the highest, CSM and EASM have the similar results. On the other hand, in terms of migration time, CSM is longest, MUSM takes the second place and EASM is the shortest.

There are several reasons to explain this result. First, CSM has the smaller migration costs due to both its closest migration strategy and less interactions between migration objects. However, it is easily to cause the immigration controller becoming a new overloaded controller because of concentrating on migration distance but ignoring controller capacity. At this time, CSM must migrate switches again, and migration time is the longest. Second, MUSM searches the controller with the maximum residual processing capacity the as immigration controller, but doesn't consider the additional network costs. Third, EASM optimizes the migration objects based on the migration efficiency, which uses the selection probability to determine the migrating switches and chooses the immigration controllers optimized by simulated annealing. Both two operations could reduce migration costs effectively. The sequential migration process of EASM also makes migration time lower.

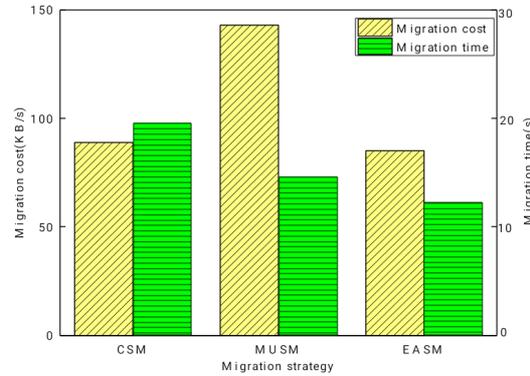

Fig. 9. Migration cost and migration time

### 6.2.4 Load balancing rate

Firstly, in order to verify the comprehensiveness of EASM strategy, we select the Internet2 OS3E as the experimental topology, and formulate a situation that there are four migrating switches with the same migration efficiency and four immigration controllers. Fig. 10 shows the load balancing rate of EASM. It is clearly seen that EASM has the higher and stable average load balancing rate, and the load balancing rate fluctuates less. This is because EASM design considers different scenarios and has the universality for the network. Especially, when switches have the same migration efficiency, EASM can still keep efficient migration according to the minimum path length among controllers. Therefore, EASM has the better general applicability for SDN network.

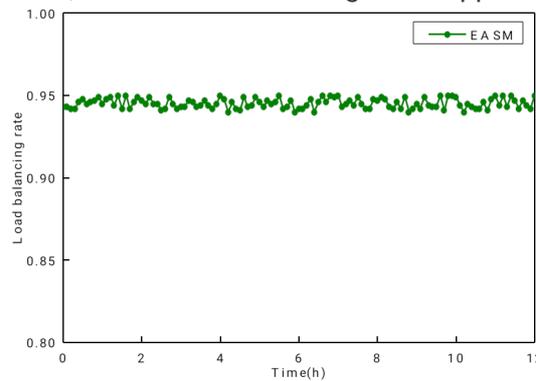

Fig. 10. Load balancing rates under multiple migrating objects with
the same migration efficiency and multiple immigration controllers

Further, we validate the load balancing performance of EASM in the other topologies selected from Topology Zoo, and the network scale gradually expands. As shown in Fig. 11, and the normalized processing is implemented for load balancing rate to make comparisons more clearly. We observe that the load balancing rate of EASM is higher than the other three strategies, and it almost does not change along with the topology expanding. This is because the setting of migration efficiency and triplet in EASM could achieve the efficient migration planning and fast switch migration. Therefore, EASM has a stronger ability to maintain the load balancing rate at a high level, and can adapt to different network topologies.

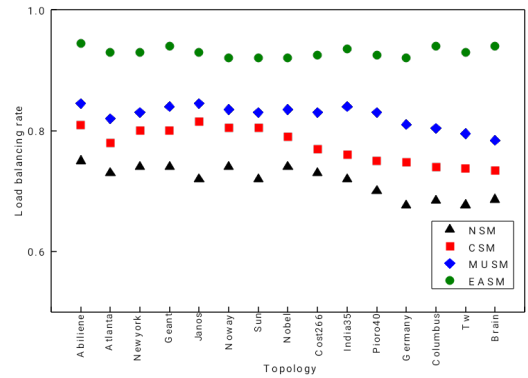
Fig. 11. Load balancing rates in different topologies

## 7. Related Work

There is a large spectrum of related work along controller load balancing. We only review some closely related ones here from the mainstream solutions.

**Controller deployment scheme:** The original SDN network relies on the centralized controller, which has the problems of low processing performance and poor scalability, so the related researchers propose to deploy multi-controller, such as HyperFlow [4], Onix [5] and Kandoo [6]. In order to balance the loads of distributed controllers, in [12], the authors firstly consider the controller deployment, and optimizes the locations of controllers based on the average delay and the maximum delay. Meanwhile, this method also introduces the deployment instances to analyze the distribution of loads. In [13], the authors design a Pareto- based Optimal COntroller (POCO) placement, which makes a compromise in controller performance, failure tolerance and load balancing. In [14], the authors propose a dynamic controller planning with load regulation, and this architecture could adjust the number of active controllers adaptively and minimize the flow setting time. However, it must collect the traffic information periodically and implement load redistribution from the entire control plane. In [15], the authors consider the usage of control resource, and design JumpFlow to reduce the usage of flow table and the ratio of average control messages. In [16], the authors propose an efficient online algorithm for dynamic SDN controller assignment, and mainly consider response time and maintenance cost. A hierarchical two-phase algorithm that integrates key concepts from both matching theory and coalitional games is designed to solve it.

**Switch migration scheme:** Following OpenFlow 1.3 protocol [10], a switch could be connected with multiple controllers in the meantime. A switch may be simultaneously connected to multiple controllers in equal state, multiple controllers in slave state, and at most one controller in master state. Each controller may communicate its role to the switch via a role request message, and the switch must remember the role of each controller connection. The subdomain controller is the master role of the subdomain switches, but those switches can set the other subdomain controllers as slave roles. Therefore, some people study controller load balancing from the perspective of switch migration. In [11], the authors design an ElastiCon architecture with double threshold values, and ElastiCon migrates switch into the closest neighbor controller. In [17], the authors propose switch migration based on clustering controller and divide the whole network into multi-domains. The dynamic allocation of controller load between multiple clusters is realized by switch migration. Meanwhile, this method also supports failover and controller backup. In [18], the switch migration is programmed as the maximum resource utilization, and the distributed hopping algorithm is designed based on Log-Sum-Exp function to approximate optimal object. Besides, it runs on each controller independently. In [19], the authors introduce load variance-based synchronization (LVS) to improve the load balancing performance in the multi-controller and multi-domain network. LVS conducts state synchronization among controllers if and only if the load of a specific server or subdomain exceeds a certain threshold. In [20], by constructing the game-playing fields, the authors design a decision-making mechanism based on zero-sum game theory to reelect a new controller as the master for the switches. In [21], the authors propose BalCon (Balanced Controller), which is an algorithmic solution designed to tackle and reduce the load imbalance among SDN controllers through proper SDN switch migrations. However, BalCon is only suitable for the small-scale network.

## 8. Conclusion

In this paper, we make the first attempt to optimize the process of switch migration through introducing the migration efficiency, and propose an Efficiency-Aware Switch Migration (EASM) strategy for balancing multi-controller loads. The essence of EASM is to migrate switches with the consideration of load balancing rate and migration costs to improve the migration efficiency and balance the controller loads. Simulation results show that EASM simultaneously achieve low controller response time, high controller throughput, low migration cost and better load balancing rate. In the future, we will improve EASM in the following aspects: (1) deploying EASM in a large-scale test bed, (2) researching the reliability of controller, (3) considering the security of the switch migration.